# Sensing Noncollinear Magnetism at the Atomic Scale Combining Magnetic Exchange and Spin-Polarized Imaging

Nadine Hauptmann, Jan W. Gerritsen, Daniel Wegner, Alexander A. Khajetoorians[*]

Institute for Molecules and Materials, Radboud University, 6525 AJ Nijmegen, Netherlands

[*] Correspondence to: a.khajetoorians@science.ru.nl

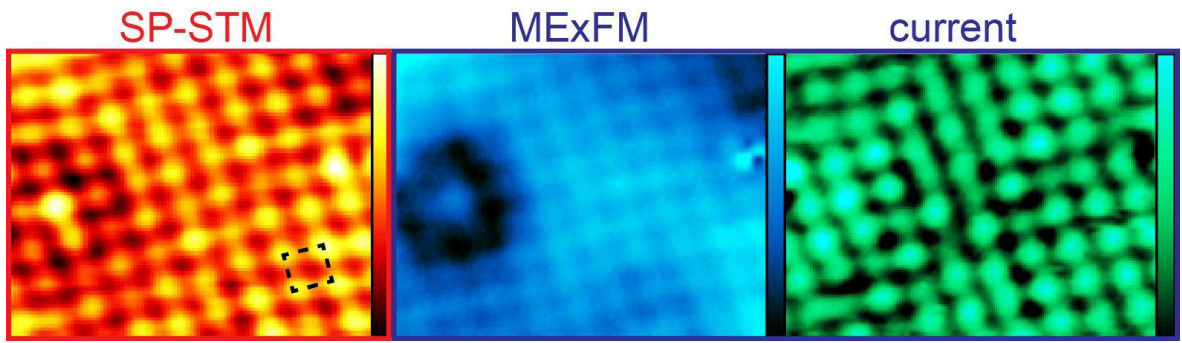

*Table of Contents Figure (TOC)*

Storing and accessing information in atomic-scale magnets requires magnetic imaging techniques with single-atom resolution. Here, we show simultaneous detection of the spin-polarization and exchange force, with or without the flow of current, with a new method, which combines scanning tunneling microscopy and non-contact atomic force microscopy. To demonstrate the application of this new method, we characterize the prototypical nano-skyrmion lattice formed on a monolayer of Fe/Ir(111). We resolve the square magnetic lattice by employing magnetic exchange force microscopy, demonstrating its applicability to noncollinear magnetic structures, for the first time. Utilizing distance-dependent force and current spectroscopy, we quantify the exchange forces in comparison to the spin-polarization. For strongly spin-polarized tips, we distinguish different signs of the exchange force which we suggest arises from a change in exchange mechanisms between the probe and a skyrmion. This new approach may enable both non-perturbative readout combined with writing by current-driven reversal of atomic-scale magnets.

The ultimate goal of magnetic-based storage is to create ultra-high density memory based on energy-efficient manipulation[1] of the remnant magnetization state of nanomagnets. Magnetic nanostructures[2-5] as well as magnetic atoms on surfaces[6-9] have emerged as candidates for atomic-scale magnetic storage. While it has recently been shown that the magnetic remanence can be greatly enhanced for a single magnetic atom on a surface[8, 9] by utilizing a combination of weakly conducting surfaces, symmetry[10], and the localized nature of 4$f$-derived moments[8], the magnetic state of such nanoscale magnets are extremely sensitive to readout techniques based on spin-polarized current[5-7, 9, 11]. To this end, various remote readout schemes[9, 12, 13] based on spin-polarized tunneling have been developed in order to probe the intrinsic magnetization dynamics of a single atom supported on thin insulating films[14, 15]. However, an atomic-scale sensing scheme, which has the freedom to operate with or without the flow of current, and which can deconvolute magnetic, electronic, and structural variations, has not been shown so far.

Spin-polarized scanning tunneling microscopy (SP-STM)[16] has emerged as the leading technique for characterizing and manipulating the magnetization of surfaces at the atomic length scale. Despite its vast success, this technique faces some limitations: (1) the topographic, electronic, and magnetic contributions are convoluted[17]; (2) sensing requires current flow, which can unintentionally flip the magnetization[6]; (3) detection is often limited to orbitals that exhibit spin polarization far into the vacuum[11]. To this end, non-contact atomic force microscopy (NC-AFM)[18] is a complementary technique to STM, providing a high-resolution method for structural and orbital characterization[19, 20] as well as chemical sensitivity[21], and operating without current flow. Based on NC-AFM, magnetic exchange force microscopy (MExFM)[22] provides an alternative means toward detecting magnetism, by directly measuring the exchange force between a magnetic probe and the sample. MExFM also provides the means of quantifying the exchange force via distance-dependent spectroscopy[23]. While SP-STM has been widely implemented, MExFM has thus far only been applied to few surfaces, with the first studies focusing on antiferromagnetic structures[22-25]. Therefore, a clear advance in magnetic imaging of nanoscale magnets would be to not only expand the application of MExFM towards noncollinear magnetic structures, but to combine SP-STM and MExFM simultaneously for a more complete picture of magnetism at the atomic scale.

Here, we demonstrate a new type of magnetic imaging based on combining both SP-STM and MExFM simultaneously, employing low-temperature STM/AFM based on a qPlus sensor[26] mounted with a ferromagnetic Fe tip, which we refer to as SPEX (spin-polarized/exchange) imaging. Using SP-STM as a starting point, we observe the well-known two-fold magnetic structure of the fcc monolayer of Fe on Ir(111)[4]. Applying height-dependent imaging, we illustrate the onset of magnetic exchange contrast of the nano-skyrmion lattice, which emerges closer to the surface as compared to typical SP-STM imaging, and we compare this to the measured spin-polarized current at that height. We observe a positive correlation between the magnetic images in both imaging modes. By employing force and current spectroscopy, we quantify the spin polarization and exchange force as a function of distance from the surface, which illustrates that substantial spin polarization exists further out in the vacuum compared to the exchange force. For all probes, the spin polarization remains nearly constant. However, for probes that exhibit a stronger spin polarization, we observe a reversal in the magnetic exchange force with increasing tip-surface separation evidencing a detectable distance-dependent transition in the exchange mechanism between the surface and probe.

Employing SP-STM in constant-current mode, we first characterize the prototype nano-skyrmion lattice of a monolayer of Fe on Ir(111)[4], which provides access to all magnetization directions. The Fe monolayer grows pseudomorphically on Ir(111) and forms two island types depending on the overall stacking of the atoms, referred to as fcc and hcp[4, 27] (Fig. 1a). The prototypical nano-skyrmion lattice is found in the fcc islands[4]. Here, the magnetic ground state is composed of Néel-type skyrmions, characterized by a square symmetry with a calculated magnetic unit cell length of $a_c \approx 1$ nm (shown in Fig. 1b for an out-of-plane spin-polarized tip)[4]. The square magnetic unit cell is superimposed onto the threefold-symmetric Fe/Ir(111) lattice, resulting in three rotational domains[4]. This nano-skyrmion lattice provides an opportunity to probe chiral magnetic structures with MExFM compared to the collinear antiferromagnets previously probed with MExFM[22-25].

In addition to the fcc Fe islands exhibiting the square-symmetric nano-skyrmion lattice, large-scale images also reveal regions with monolayer hcp islands as well as bilayer Fe islands (Fig. 1a). Our SP-STM images allow for distinguishing the hcp islands from their fcc counterparts because of a threefold-symmetric spin contrast, as expected[27]. Likewise, we detect the spin contrast on bilayer islands, which exhibit a complex spin-spiral network[17]. For the latter, three rotational domains coexist, which allows for calibrating the tip magnetization (Supplementary Section S2). For all measurements shown here, we ensured that the tip dominantly exhibits an out-of-plane magnetization relative to the surface, although a small in-plane component cannot be ruled out.

After characterizing the sample using SP-STM, we subsequently perform AFM in constant frequency shift mode, or so-called AFM topography. As the frequency shift $\Delta f$ is related to the average force gradient[18], the AFM topography is sensitive to variations in short-range forces here. At larger tip-sample separations, the exchange interaction and other magnetic forces are negligible, as well as spatial variations arising from electrostatic and van der Waals interactions. Therefore, the AFM topography primarily reflects the structural variations on the surface, while SP-STM topography is sensitive to local changes in the electronic and magnetic structure. Therefore, the comparison of AFM and STM images offers the possibility to distinguish between structural and electronic effects. For the acquisition of an AFM topography, we move the tip closer to the surface to an offset position $z = -0.22 \pm 0.01$ nm, where $z = 0$ nm is defined by the SP-STM stabilization parameters ($V_S = 50$ mV, $I_t = 100$ pA), and a negative value refers to a displacement toward the surface. We record the AFM topography at constant frequency shift (using $z_{\text{mod}} \leq 110$ pm) and simultaneously measure the spin-polarized current that results from a very small sample voltage ($|V_S| < 1.7$ mV). Various features are identified that can be attributed to adsorbates or structural defects in the Fe layer or beneath the surface (presumably in the iridium), which appear quite differently in the AFM vs. STM images. For example, the AFM topography clearly reveals subsurface defects both underneath fcc and hcp islands (see arrows in Fig. 1c). In the STM image (Fig. 1a), these defects cannot be easily seen on the fcc islands, while they show up as pronounced depressions on hcp islands. This is also found in the spin-polarized current map (Fig. 1d) simultaneously acquired with the AFM image.

In order to reveal the short-range magnetic exchange interaction in AFM topography, we have to move the tip closer to the surface. In the following, we focus on the fcc monolayer Fe islands exhibiting the square nano-skyrmion lattice (Fig. 2). Again, we first take an SP-STM image using the above-mentioned stabilization parameters that define $z = 0$ nm (Fig. 2a). For the MExFM imaging, we change the tip offset roughly to $z = -0.31$ nm and take a constant frequency shift image (b) while simultaneously acquiring a current map (c). The constant frequency shift image reveals a square structure (Fig. 2b) with a lattice constant $a = 0.87 \pm 0.4$ nm, which is comparable to the values taken from the SP-STM images of the nano-skyrmion lattice (Figs. 1a and 2a) and in reasonable agreement with previously reported experimental values[4]. An identical pattern is found in the simultaneously acquired current map (Fig. 2c), resulting from the retained spin polarization of the current at this height. Line profiles at identical positions of both images reveal the clear correlation of both signals (Fig. 2d). Therefore, we conclude that the constant frequency shift image corresponds to a MExFM image, i.e., we detect the spatial variation in exchange force between tip and surface. We note that the tip magnetization is constant, in other words we see no evidence that the tip magnetization flips as a function of distance or due to superparamagnetic fluctuations. Therefore, the image contrast corresponds to the difference in the attractive force between the aligned and anti-aligned orientation of the magnetic moments within the nano-skyrmion lattice relative to the tip moment. We discuss this in more detail below.

A quantitative comparison of the line profiles in Fig. 2d reveals that the MExFM contrast between aligned and anti-aligned portions of the nano-skyrmion lattice are typically in the range of $\Delta z = 1.0 \pm 0.5$ pm. As we detail below, the magnetic contrast corresponds to a frequency-shift difference of only $\Delta f \approx 0.1$ Hz. This corrugation in the MExFM images is about a factor 10 smaller than reported for the antiferromagnetic Fe monolayer on W(001)[25], which suggests that the overall exchange force between the ferromagnetic tip and the nano-skyrmion lattice may be weaker for this system. We observe that a larger height corrugation in the MExFM images is correlated with a larger current.

In combined STM/AFM imaging with a tuning fork, it is important to rule out potential cross-talk between the current and the frequency shift, where the tunneling current may introduce an interference with the deflection of the tuning fork. In order to account for that, we acquired constant height images of the frequency shift (Fig. 3(a)) and the simultaneously measured current (Fig. 3(b)). The resultant images both reveal the aforementioned skyrmion lattice. The constant height image qualitatively reproduces the contrast variations seen in constant frequency-shift imaging. In order to exclude cross talk, we changed the applied sample voltage during image acquisition, with feedback off (arrows). While the change in voltage leads to a change in current, leading to a strong contrast variation, the measured frequency shift is not influenced by this strong perturbation (see also Supplementary Section S8). This also holds for MExFM images (constant frequency shift mode, Fig. 3(c)) and the simultaneously measured current in (d), allowing us to ascertain that both the frequency shift and current channels are independent.

To quantify the exchange force and spin polarization as a function of tip-surface separation, we perform distance-dependent measurements by moving the tip toward the surface and simultaneously recording the variations in frequency shift (using $z_{mod}$ = 40-80 pm) and current for dozens of different out-of-plane magnetized tips (representative curves are shown in Fig. 4; Supplementary Section S7 for all data and the acquisition procedure). We perform these measurements at positions of maximum contrast, corresponding to aligned/anti-aligned magnetization orientation relative to the tip magnetization[4]. In order to exclude effects from a spatially dependent difference in tip height, the current feedback was opened prior to the movement towards the surface (SP-STM stabilization parameters $V_S$ = 50 mV, $I_t$ = 100 pA) and the tip was moved to the different lateral positions in constant height mode (see also Supplementary Section S7). The variations in current as a function of distance exhibit the expected exponential dependence (Fig. 4b), with the spin polarization between aligned and anti-aligned orientations varying slightly at all probed displacements. The negative frequency shift increases as the tip-sample separation is decreased (Fig. 4a), indicating a stronger attractive force between the tip and the surface. At first glance, there is no obvious difference between the frequency-shift curves for aligned vs. anti-aligned orientations on the nano-skyrmion lattice. However, as discussed above, the MExFM contrast is very small, while the overall $\Delta f$ is a sum of the chemical, electrostatic, magnetic dipole and exchange forces. Therefore, we revert to a previously applied method to extract magnetic exchange forces where the frequency shift due to antiferromagnetic exchange interaction has been defined by the difference between the bright and dark positions in the AFM topography, corresponding to anti-aligned and aligned magnetic moments between the iron atoms and the tip, respectively[23]. Based on this, we examine the difference between the bright ($\Delta f_\uparrow(z)$) and dark ($\Delta f_\downarrow(z)$) positions of the AFM topography, i.e., $\Delta f_{ex}(z) = \Delta f_\uparrow(z) - \Delta f_\downarrow(z)$, in order to quantify the magnetic exchange interaction between the different out-of-plane magnetization directions of the nano-skyrmion lattice. Fig. 4c shows representative difference curves for three different tips. They reveal that $\Delta f_{ex}(z)$ starts to decrease at $z \approx -0.2$ nm, which indicates the onset of a significant exchange force between tip and surface. The frequency shift related to the magnetic exchange is on the order of 0.1 Hz, i.e., about a factor of 400 smaller than the overall measured frequency shift (Fig. 4a). We note that distance-dependent data acquired in between the positions of out-of-plane magnetization directions does not show any evidence of noncollinearity.

The distance-dependent curves also permit to compare exchange-force frequency shifts $\Delta f_{ex}$ with the spin polarization. For this, we plot the current asymmetry $\mathcal{A}(z) = (I_\uparrow(z) - I_\downarrow(z))/(I_\uparrow(z) + I_\downarrow(z))$ as a function of $z$ (Fig. 4d), which was simultaneously acquired with the $\Delta f_{ex}$ curves using the same tips (as color-coded in Figs. 4c,d). The plots show that $\mathcal{A}(z)$ increases only slightly as the tip height is decreased, indicating that the spin polarization does not reverse, and remains relatively constant. The small variation and constant sign of $\mathcal{A}(z)$ is further evidence that the tip magnetization remains nearly constant within the entire probed regime, allowing us to rule out that the increasing exchange interaction between tip and surface strongly affects the tip magnetization. The comparison of $\Delta f_{ex}(z)$ with $\mathcal{A}(z)$ also illustrates the different height regimes at which SP-STM and MExFM work, illuminating the complementary information that can be acquired by both techniques.

Finally, to extract the magnetic exchange force $F_{ex}(z)$ from $\Delta f_{ex}(z)$, we utilize the formula[28]

$$F_{ex}(z) = 2k \int_z^\infty \left(1 + \frac{z_{mod}^{1/2}}{8\sqrt{\pi(t-z)}}\right) \frac{\Delta f_{ex}(t)}{f_0} - \frac{z_{mod}^{3/2}}{\sqrt{2(t-z)}} \frac{d}{dt}\left(\frac{\Delta f_{ex}(t)}{f_0}\right) dt \quad [1]$$

with $f_0$ and $z_{mod}$ being the resonance frequency and oscillation amplitude, respectively. The stiffness of the tuning fork ($k \approx 1800$ N/m) was taken from literature[26]. The resultant is plotted in Fig. 4e. The onset of the exchange force can be clearly seen at displacements z ≈ −0.2 nm. All $F_{ex}(z)$ curves show a decrease of the force, down to −25 pN.

In addition to observing an onset of a strong exchange force (z ≈ −0.4 nm), resulting from direct exchange between Fe atoms, we see a significant exchange interaction of opposite sign character at larger separations for certain tips (z ≈ −0.3 nm). By cross-correlating the extracted exchange force $F_{ex}(z)$ with the spin-polarization $\mathcal{A}(z)$, our data suggests that this force reversal is particularly evident for tips that exhibit a larger spin polarization $\mathcal{A} > 0.3$ (Fig. 4d). Now we discuss the various mechanisms that can modify the exchange interaction between the probe and surface. In refs[29, 30], it was shown that relaxation of the foremost atoms of a Cr tip can induce a modification in the exchange interaction. However, in that case the effect was strongest for Cr tips on Fe surfaces, whereas we have purely Fe probes here. Nevertheless, a sign reversal in the exchange energy was calculated and assigned to an indirect exchange mechanism[23, 30] when the Cr tip was brought closer to the antiferromagnetic Fe layer on W(100). For that system, the probe can sense the exchange from additional neighboring atoms in addition to the atom beneath the tip at certain separations. The higher number of anti-aligned Fe atoms, compared to aligned Fe atom underneath the tip leads to an antiferromagnetic exchange between the probe and surface[23]. However, for the Fe skyrmion lattice studied here, the magnetization of nearest-neighbored atoms are only partially rotated with respect to the magnetization of the atom directly underneath the tip, due to the small pitch change in the skyrmion unit cell. Therefore, an antiferromagnetic interaction of neighboring Fe atoms on the surface is considered unlikely.

The stray field of the Fe tip may also modify the measured forces. As the spin-polarization of our probes is constant over the accessed distance range (Fig. 4d), a stray field from the tip will add an offset and alone cannot change sign of the exchange force at larger separations. Finally, in the calculations in ref. [31] for a single magnetic atom on top of a metallic surface, it is shown for certain probes that a sign change in the interaction can be observed as a function of probe-surface separation.

In this case, the sign reversal is attributed to a change in direct overlap of *s* or *p* orbitals with the *d* orbitals, i.e. a Zener model. We speculate that this last mechanism may be responsible for the observed sign reversal. However, we note that they predict a large change in the spin-polarization as a function of distance, which is inconsistent with our observations. We also note, however, that also a few tips with low spin polarization (Supplementary Section S7) exhibit a reversal and a few tips with large $\mathcal{A}$ do not, indicating that other effects, e.g. the tip-apex geometry or relaxations, may also play a non-negligible role. The alignment of the tip magnetic moment with respect to the local moments of the skyrmion lattice, i.e. the assignment of ferromagnetic/antiferromagnetic interactions, requires comparison to *ab-initio* calculations in order to determine the relative orientations responsible for the imaged intensity variation in the magnetic unit cell[23, 29-32].

We employed a new method of magnetic imaging, by combining SP-STM and MExFM based on STM/NC-AFM with a tuning fork. We illustrate that this new combination can be utilized to characterize chiral magnetic structures, as exemplified by the nano-skyrmion lattice. SPEX imaging could provide complementary information by deconvoluting structural features from electronic and magnetic properties, which is typically very difficult to decouple. For example, we see evidence in the bilayer of Fe/Ir(111) of strong vertical relaxations resulting in non-planar structures in AFM imaging (Fig. 1c) which may be related to the dislocation network which was previously reported in pure SP-STM imaging[17, 33]. The combined method can provide more complete characterization toward understanding the impact of defects on the magnetic ground state, as well as a path toward studying multi-element magnetic surfaces that can be difficult to delineate based on STM alone. Distance-dependent spectroscopy reveals the different height regimes at which spin polarization and various types of magnetic exchange can be detected above the surface. While the spin polarization is nearly constant for a given probe, we observed that the magnetic forces at large separations depend strongly on the absolute spin polarization of the tip. To this end, combining SPEX imaging with *ab-initio* methods would be advantageous in revealing the relevant exchange mechanisms and surface atoms responsible for the measured force, and correlating this with the spin polarization of the tip[32]. Moreover, investigation of other noncollinear surfaces would be interesting, in order to ascertain if the MExFM method can detect noncollinear exchange. The advantages of probing magnetism at closer distances compared to spin-polarized tunneling may also enable direct access to the strongly localized and elusive 4$f$ orbitals[8, 9] in future experiments, where tunneling-based experiments have been inconclusive or could only indirectly probe 4$f$ magnetism[9, 11]. After preparation and submission of this manuscript, we became aware of similar work utilizing solely MExFM[34].

**Methods**

Scanning probe microscopy was performed utilizing a commercial ultra-high vacuum low-temperature STM/AFM from CreaTec Fischer & Co GmbH, which operates at a base temperature of $T = 6.3$ K. Atomic force microscopy (AFM) measurements using a non-contact frequency-modulation mode were done utilizing a tuning fork-based qPlus sensor[26] with its free prong oscillating at its resonance frequency $f_0 \approx 27.7$ kHz. The force is indirectly measured by the shift of the resonance frequency $\Delta f$. Oscillation amplitudes $z_{mod}$ between 40 pm and 110 pm were used with $z_{mod}$ being half the peak-to-peak value. As we do not observe a minimum in $\Delta f$, all data is acquired in the attractive force regime. Further details on the experimental parameters, tip variations and absence of crosstalk are available in Supplementary Sections S1, S6, and S8, respectively.

The Ir(111) surface was prepared by repeated cycles of Ne$^+$ sputtering and annealing ($T \sim 1800$ K) in an oxygen atmosphere ($p \sim 4 \times 10^{-6}$ mbar) followed by a final flash to 1800 K. The Fe was deposited from an e-beam evaporator onto Ir(111) kept at room temperature and subsequently annealed ($T \sim 630$ K), leading to the formation of multilayer Fe islands, in which the first layer exhibits both hcp and fcc stacking.

**Acknowledgements**

We would like to acknowledge financial support from the Emmy Noether Program (KH324/1-1) via the Deutsche Forschungsgemeinschaft, and the Foundation of Fundamental Research on Matter (FOM), which is part of the Netherlands Organization for Scientific Research (NWO), and the VIDI project: 'Manipulating the interplay between superconductivity and chiral magnetism at the single atom level' with project number 680-47-534 which is financed by NWO. NH and AAK also acknowledge support from the Alexander von Humboldt Foundation via the Feodor Lynen Research Fellowship. We would like to thank A. K. Lemmens for technical support during the experiments.

**Supporting Information Available**

The Supporting Information is available free of charge on the ACS Publications website. It contains further details about the experimental methods, tip characterization, the amplitude and excitation signals, as well as on data acquisition procedures.

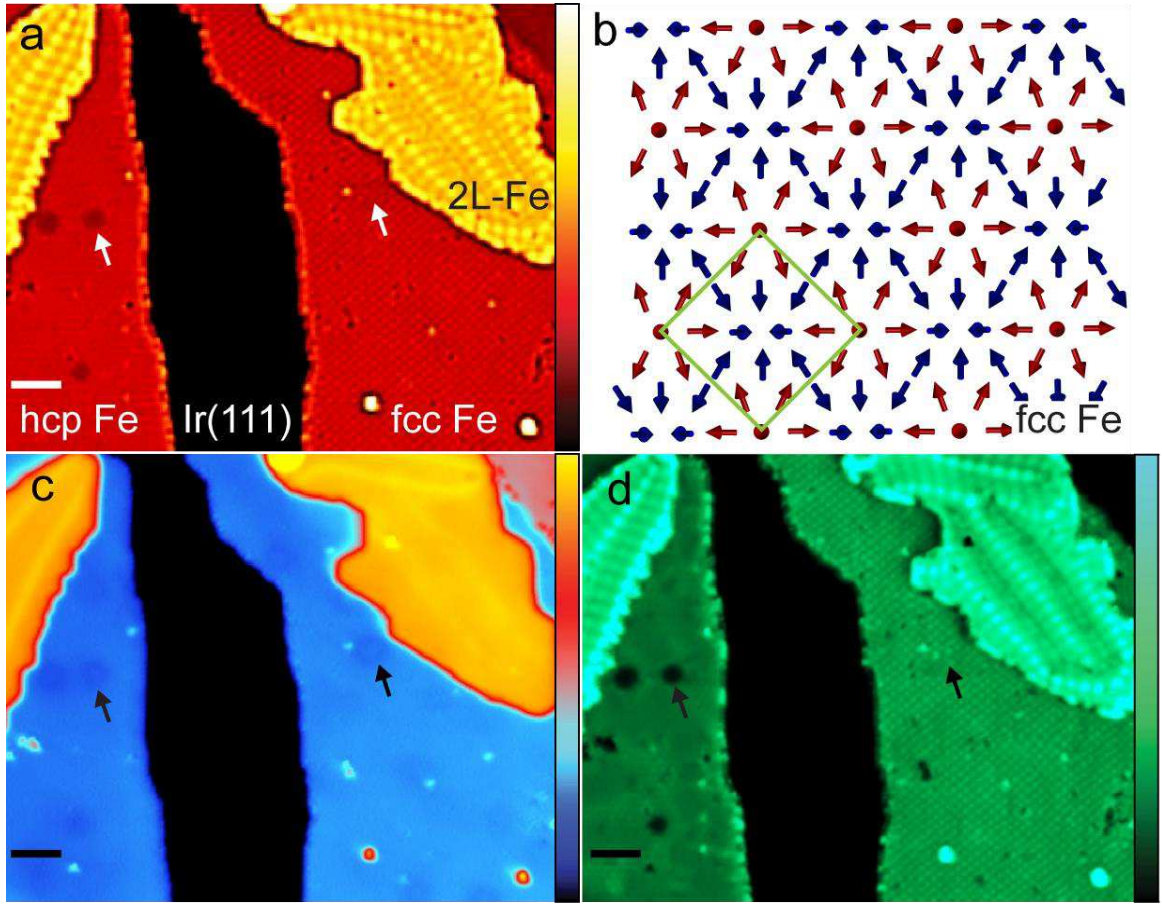

**Figure 1:** (a) Large-scale mapping (55 x 45 nm$^2$) of mono- and bilayer islands of Fe on Ir(111) for SP-STM ($V_S$ = 50 mV, $I_t$ = 100 pA), merged with the Laplace-filtered image to highlight structural details (for raw data see Supplementary Section S3). The labels hcp/fcc refer to the stacking of the monolayer islands, where fcc exhibits the square-symmetric magnetic nano-skyrmion structure, and 2L refers to the bilayer, where complex spin spirals can be observed. (b) Schematic atomic-scale view of the nano-skyrmion lattice. For the sake of clarity the commensurate representation is shown. (c) Constant frequency shift AFM topography revealing structural details of the surface such as adsorbates and subsurface defects ($\Delta f_{set} = -10.9$ Hz, $z_{mod} = 64$ pm, $V_s = 0.6$ mV, $z = -0.22 \pm 0.01$ nm relative to the image in (a)). (d) Tunneling current map simultaneously recorded with the AFM topography in (c), which includes the spin-polarized signal for each layer. The lateral scale bars in all panels correspond to 5 nm. Color-scale ranges: (a) 0.33 nm to 1.1 nm, (c) 0.1 nm to 0.42 nm (d) 3.5 pA to 213 pA.

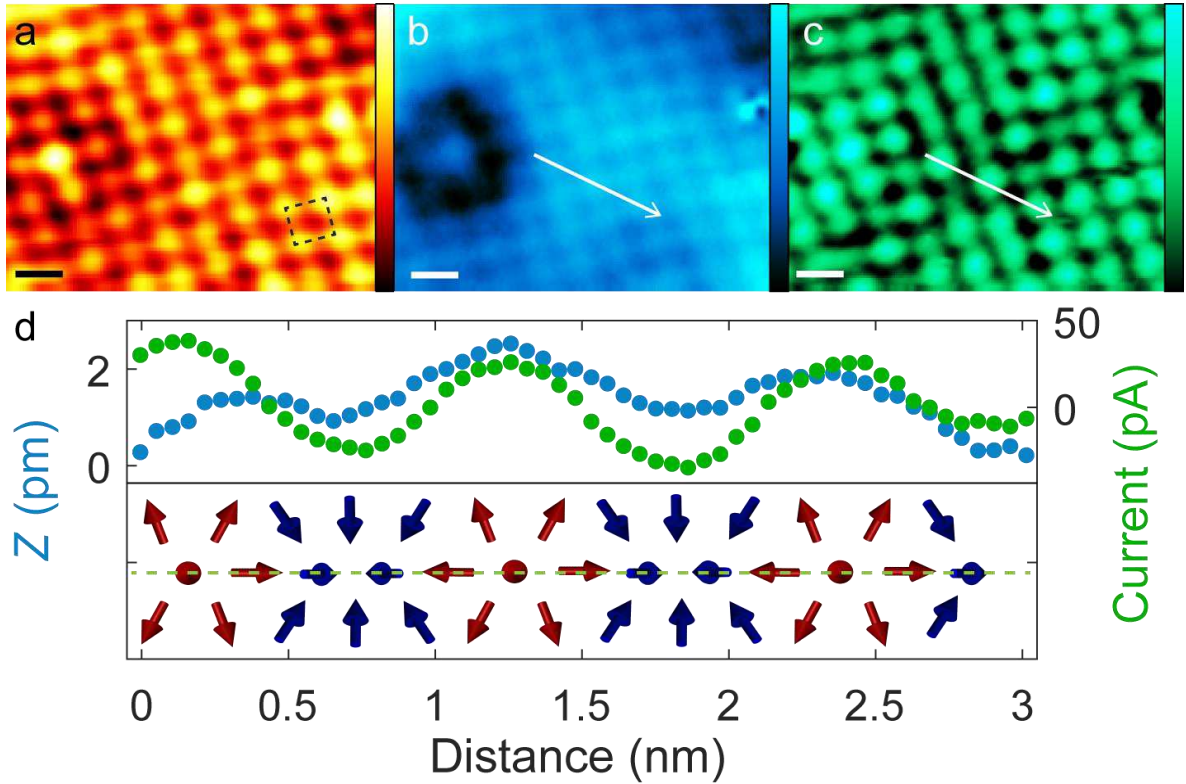

**Figure 2:** High-resolution mapping (8 x 6 nm$^2$) of the nano-skyrmion lattice (a) SP-STM ($V_S$ = 50 mV, $I_t$ = 100 pA). The dashed square indicates the magnetic unit cell. (b) MExFM image in constant frequency shift mode ($\Delta f_{set}$ = −14.6 Hz, $z_{mod}$ = 102 pm, $V_S$ = 0.2 mV, $z$ = −0.31 ± 0.01 nm). (c) Simultaneously acquired current map. (d) Line profiles along the arrows indicated in (b) and (c). A simplified view of the magnetization of the nano-skyrmion lattice is shown below the line profile. The scale bar in all panels is 1 nm. The color-scale ranges are: (a) −10.2 pm to 12.2 pm, (b) 0.11 pm to 9.3 pm (c) −26 pA to 48 pA.

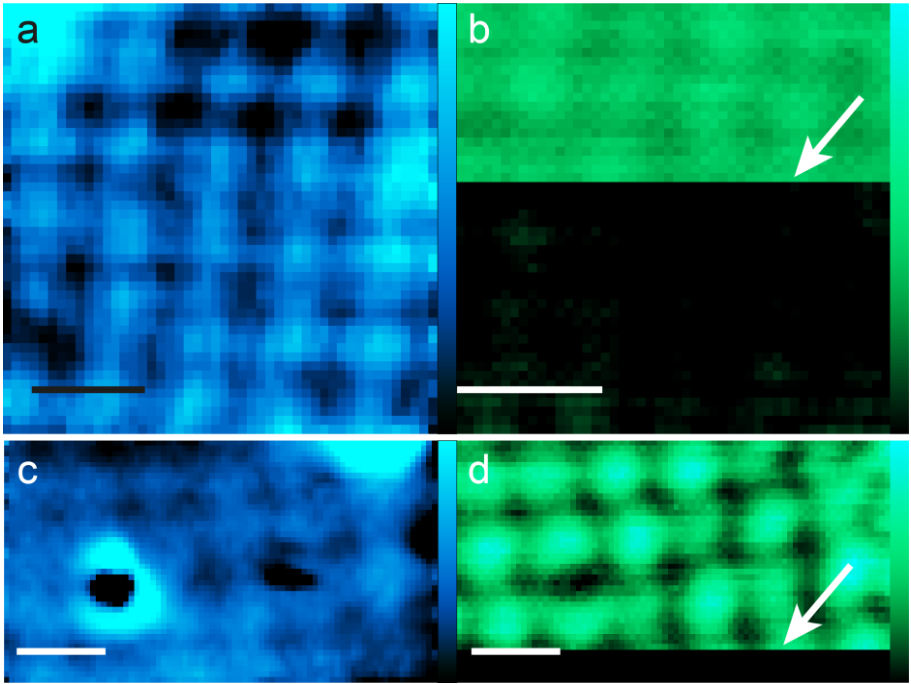

**Figure 3:** (a) Frequency shift (smoothed by a Gaussian filter, see Supplementary Section S8 for raw data) and (b) simultaneously current map acquired in constant height mode ($z_{mod}$ = 51 pm, $z$ = –400 pm with the current-feedback loop opened above the bright position of the skyrmion lattice at $V_s$ = 50 mV and $I_t$ = 100 pA). At the line indicated by the arrow, the bias voltage was changed from $V_S$ = 0.1 mV (upper part) to 0 mV (lower part). (c) MExFM image (smoothed by a 2-point Gaussian filter, see Supplementary Section S8 for raw data) and (d) simultaneously measured current map in constant frequency shift mode ($\Delta f_{set}$ = –36 Hz, $V_s$ = 0.0 mV (lower part), $V_s$ = 0.4 mV (upper part), $z_{mod}$ = 180 pm)). At the scan line marked by the white arrow the voltage was increased from 0 to 0.4 mV. The scale bar is 1 nm in all images. Color scales: (a) –34.5 to –34.0 Hz, (b) –214 pA to 340 pA, (c) 14.5 to 17.8 pm, (d) 0.9 to 1.8 nA.

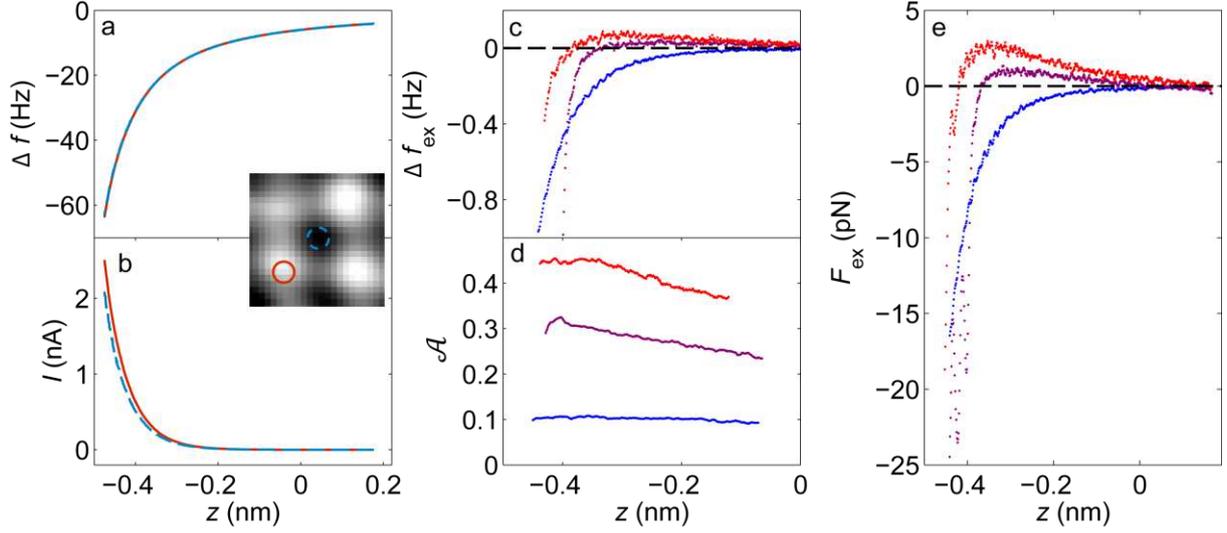

**Figure 4:** (a) Distance dependence of $\Delta f(z)$ at two different locations with opposite contrast within the magnetic unit cell of the nano-skyrmion lattice (cf. areas circled in red ($\Delta f_\uparrow(z)$) and blue ($\Delta f_\downarrow(z)$) in the inset), acquired with an dominantly out-of-plane magnetized tip ($z = 0$ nm is defined by STM stabilization parameters $V_S = 50$ mV, $I_t = 100$ pA). (b) Simultaneously acquired distance dependence $I(z)$ at the same two locations using the same tip. (c) Difference in frequency shift $\Delta f_{ex}(z) = \Delta f_\uparrow(z) - \Delta f_\downarrow(z)$, revealing the magnetic exchange contribution, for three different tips (Savitzky-Golay filtered prior to subtraction). (d) Distance-dependent spin-polarized asymmetry ($\mathcal{A}(z) = (I_\uparrow(z) - I_\downarrow(z))/(I_\uparrow(z) + I_\downarrow(z))$) for the same three tips used in (c) (cf. color code). (e) Exchange force $F_{ex}(z)$ extracted from $\Delta f_{ex}(z)$ using Eq. 1 (prior to this, $\Delta f_{ex}(z)$ was smoothed using a Savitzky-Golay filter). The color code for (c) to (e) reflects low (blue) to high (red) spin asymmetry $\mathcal{A}(z)$. The supplementary information provides distance dependences for dozens of different tips (section S7) to reflect the reproducibility and statistical spread.

# Supporting Information:

# Sensing Noncollinear Magnetism at the Atomic Scale Combining Magnetic Exchange and Spin-Polarized Imaging

Nadine Hauptmann, Jan W. Gerritsen, Daniel Wegner, Alexander A. Khajetoorians[*]

Institute for Molecules and Materials, Radboud University, 6525 AJ Nijmegen, Netherlands

[*] Correspondence to: a.khajetoorians@science.ru.nl

**S1. Experimental details**

All measurements were performed with an Fe bulk tip attached to the free prong of the tuning fork. Two different types of tuning forks were used: (a) an asymmetric tuning fork with the Fe tip contacted via an additional electrode printed on the prong; (b) a symmetric tuning fork where the Fe tip is contacted by a 12 µm thin wire. The latter solution is suggested to avoid cross talk between the tunneling current and the deflection of the tuning-fork prong.[1,2] The tuning-fork oscillation amplitude was calibrated using the tunneling current-controlled amplitude determination[3]. Owing to the calibration procedure, an error of the oscillation amplitude of 10% is assumed.

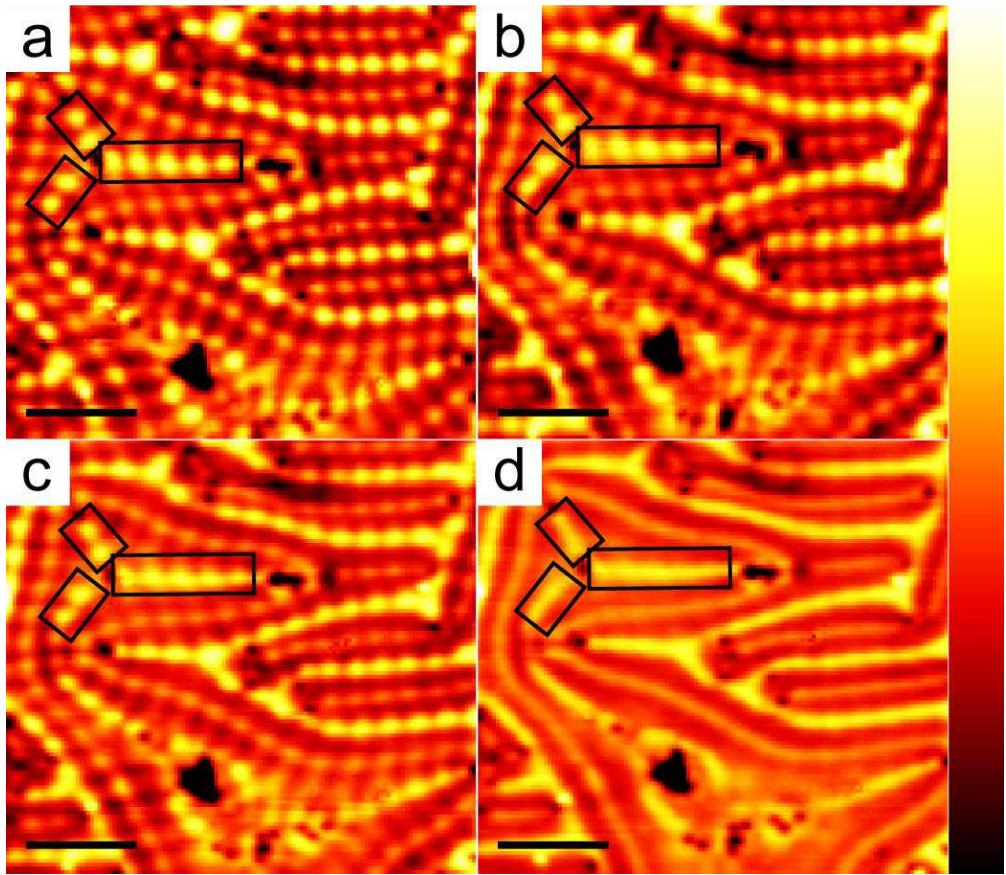

**Figure S1: Characterization of the tip magnetization:** SP-STM constant-current topography (raw data) of different domains on a bilayer Fe island (25 x 25 nm$^2$, $V_s$ =50 mV, $I_t$ = 291 pA) for (a) a tip with a dominating out-of-plane magnetization, (b) and (c) tips with out-of-plane as well as in-plane magnetization, (d) a non-magnetic tip. The three boxes in every image indicate the apparent-height modulation along all three rotational domains of the spin spiral. The scale bar is 6 nm in all images. Color scales: (a) 0.19-0.27 nm, (b) 0.16-0.23 nm, (c) 0.10-0.19 nm, (d) 0.17-0.28 nm.

The constant frequency shift images were acquired in the attractive force regime. While for larger tip-surface separations ($d \gtrsim 500$ pm) long-range van der Waals, electrostatic and magnetic dipolar forces dominate the total tip-sample force, short-range forces become dominant at smaller separations $d \approx 100$ pm.[4] Therefore, the oscillation amplitude was chosen such that it is in the order of the range of the tip-surface interaction, minimizing the amplitude noise and enhancing the sensitivity to short-range interactions.[4, 5] Thus, we used oscillation amplitudes $z_{\text{mod}} \leq 110$ pm.

### S2. Tip preparation and characterization

In order obtain spin contrast, the Fe tip attached to the tuning fork is dipped into regions where the Fe islands on Ir(111) are 3-4 monolayers thick. To check the orientation of the tip spin polarization, we image bilayer Fe islands and compare the spin contrast within each of the spin-spiral domains that are rotated by 120° with respect to each other (highlighted by boxes in Figs. S1a-d). The modulations in the apparent height along the spin spirals have a spacing of about 1.4 nm. For tips with a dominating out-of-plane magnetization, the SP-STM contrast along the spin spiral is identical on all three rotational domains[6] (Fig. S1a), while a tip with significant in-plane magnetization results in different amplitudes along the three different directions or smaller total spin contrast (Fig. S1b and c). Tips with no spin polarization lead to images where the spin spirals appear blurred[6] (Fig. S1d).

In order to account for small long-range van der Waals interactions, we mainly use tips that exhibit a frequency shift $|\Delta f| < 7$ Hz at typical tunneling parameters (50 mV, 100 pA).

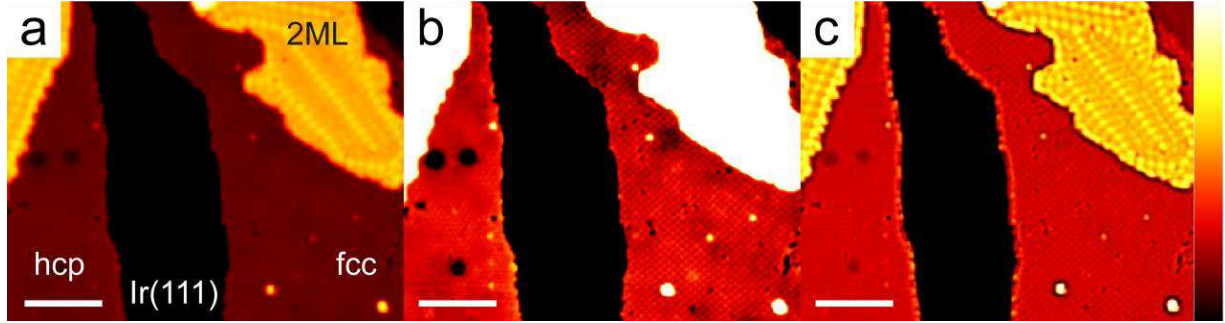

**Figure S2: Image processing of Fig. 1:** (a) and (b) shows raw-data images of the SP-STM constant-current topography shown in Fig. 1 of the main text with two differently adjusted color scales in order to see (a) the contrast of the Fe bilayer spin spirals and (b) the defects and nano-skyrmion lattice in the Fe monolayer. (c) shows the images merged with a Laplace-filtered image (as shown in the manuscript), thus enabling visualization of all the structures within a single image (55 x 45 nm$^2$, $V_s$ =50 mV, $I_t$ = 100 pA). The scale bar is 10 nm. Color scales: (a) 0.25-0.82 nm, (b) 0.30-0.38 nm, (c) 0.33-1.1 nm.

### S3. Raw-data image of Figure 1

The SP-STM image shown in Fig. 1a was processed in order to visualize both the square nano-skyrmion lattice in the fcc monolayer as well as the spin spirals in the bilayer Fe islands. For that purpose, the image was merged with a Laplace-filtered image. Fig. S2 shows the raw data images again with the color contrast adjusted to (a) the bilayer spin spiral, (b) the monolayer fcc nano-skyrmion structure. (c) shows the processed image again as shown in Fig. 1a.

### S4. MExFM-SP-STM with different defects/ adsorbates on the fcc skyrmion lattice on the Fe monolayer on Ir(111)

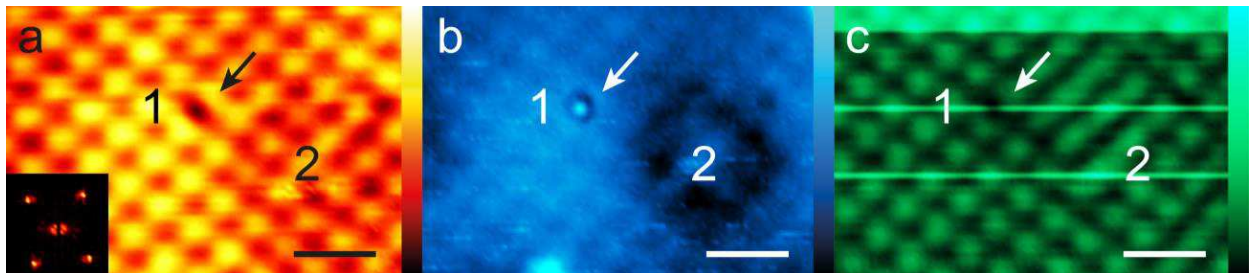

**Figure S3: MExFM-SP-STM with different defects:** (a) SP-STM constant-current topography (8.3 x 5.9 nm$^2$, $V_s$ = 50 mV, $I_t$ = 100 pA) of a fcc Fe island showing the square skyrmion lattice and two different types of defects, labeled with 1 and 2. The inset shows the 2D FFT of the image showing the square structure of the fcc skyrmion lattice. (b) MExFM image of the image in (a). (c) Simultaneously measured current map. (b) and (c): $\Delta f_{set} = -39$ Hz, $V_s = 0.1$ mV, $z_{mod} = 77$ pm. The scale bar is 2 nm in all images. Color scales: (a) 0.2-35 pm, (b) 0-16 pm, (c) 0.1-1.1 nA.

Fig. S3a shows a SP-STM image of a region with two defects, labeled "1" and "2." While defect 1 appears as a depletion in the apparent height and is spatially confined, defect 2 only leads to a small height variation. The MExFM image in Fig. S3b and the simultaneously measured current map in Fig. S3c give further information. The MExFM image shows that defect 1 is indeed an adsorbate on the fcc skyrmion lattice, while defect 2 also results in a height corrugation, but is less spatially defined. We suggest that defect 2 might be a defect in the Ir or an adsorbate in between the Ir substrate and the Fe layer. The bright lines in Fig. S3c show changes of the tip magnetization that influence the spin-polarized current, but not the MExFM contrast in Fig. S3b.

### S5. Amplitude, excitation voltage, and frequency shift for the MExFM image in Fig. 2

In order to give further information on the measurement parameters and mechanisms of Fig. 2b and c, the simultaneously measured amplitude, excitation voltage and frequency shift are shown in Fig. S4. The amplitude and the frequency shift appear without any features indicating no disturbances during the measurements. In particular, the featureless image of the excitation voltage (b) indicates a non-dissipative interaction.

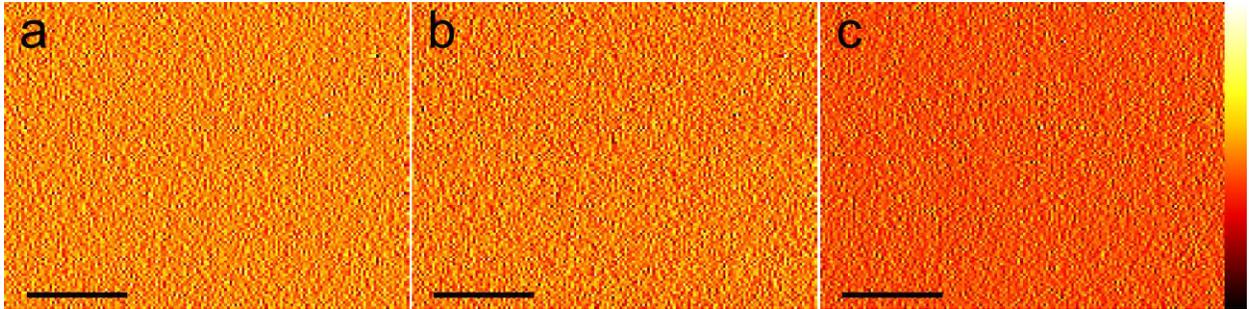

**Figure S4: Additional channels for Fig. 2b and c:** (a) Amplitude, (b) excitation voltage of the tuning fork, and (c) frequency shift $\Delta f$ simultaneously measured with the MExFM image in Fig. 2b ($\Delta f_{set} = -14.6$ Hz, $z_{mod} = 102$ pm, $V_S = 0.2$ mV). The scale bar is 2 nm in all images. Color scales: (a) 102-102.8 pm, (b) 1.74-3.93 mV, (c) −14.75 to −14.58 Hz. A roughness analysis distribution for the images yields a Gaussian distribution with a FWHM of (a) 0.24 pm, (b) 0.65 mV, and (c) 43 mHz.

### S6. MExFM data with shifted contrast in the MExFM and current image

The images shown in Fig. S5 were taken with a tip that had a frequency shift offset of −10 Hz, indicating a larger contribution from long-range van der Waals forces indicating a blunt tip. In this case, we observed a lateral shift of the corrugation between the MExFM image (Fig. S5b) and the simultaneously measured current map (Fig. S5c), as indicated by the lines that mark the same position. However, for most of our tips (10 out of 12 tips with which we acquired MExFM images) the corrugation maxima in the MExFM image and in the spin-polarized current map are at the same positions. We believe that this shift can be caused by modifications of the spin polarization relative to the exchange force for tips that

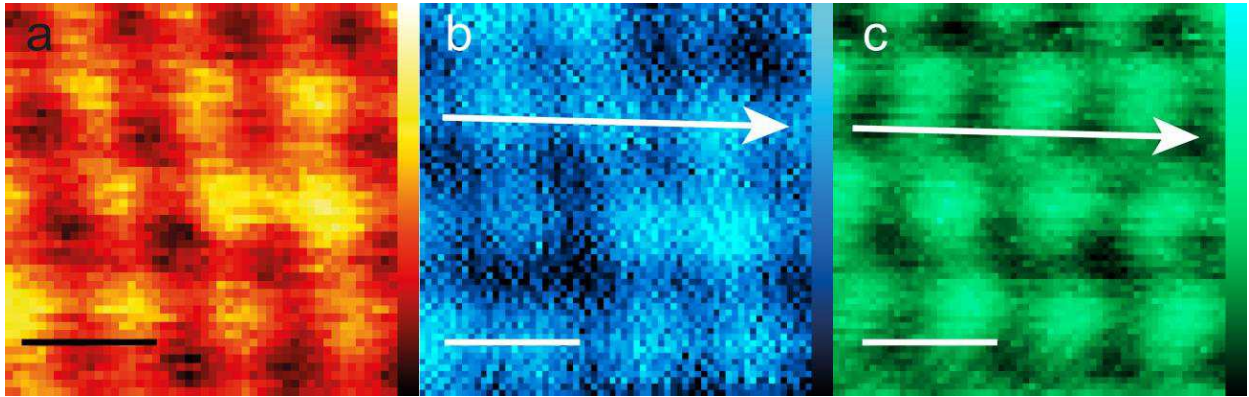

**Figure S5: MExFM for a tip leading to a shifted contrast:** (a) SP-STM constant-current topography ($V_s$ = 50 mV, $I_t$ = 100 pA), (b) MExFM image ($\Delta f_{set}$ : –60 Hz, $V_s$ = 0.2 mV, $z_{mod}$ = 58 pm). The frequency shift offset at $z = 0$ is –10 Hz indicating strong long-range van der Waals forces caused by a blunt tip. The arrows in (b) and (c) show the shift of the corrugation between the AFM topography and the simultaneously measured current. All image sizes: 3 x 3 nm². The scale bar is 1 nm in all images. Color scales: (a) 0-20 pm, (b) 1.3-4.5 pm, (c) 852 pA-1.5 nA.

change their structure near the apex.[7]

### S7. Procedure for acquisition of frequency shift vs. distance curves

The tip prepared as described in section 6 was stabilized above the position of the skyrmion lattice with maximum spin-polarized signal (cf. red-circled area of the inset image in Fig. 4). The tuning fork oscillation was then started with oscillation amplitudes (defined as half the peak-to-peak value) between 40 pm and 80 pm. The current-feedback loop was then switched off at $V_s$ = 50 mV and $I_t$ = 100 pA. The voltage was decreased to values between 0.1 mV and 0.3 mV, the current-voltage preamplifier gain is set between $10^7$ - $10^9$.

Next, the tip was retreated to $z = 200$ pm in order to probe the long-range forces and then approached to the surface to values down to $z = −480$ pm. We measured both the approach and retract sweeps and recorded the frequency shift, spin-polarized current, amplitude, and excitation voltage. Two to eight sweeps were acquired at this position and averaged before going to the neighboring position of opposite spin contrast within the skyrmion lattice (cf. blue-circled area of the inset image in Fig. 4), with the current-feedback loop switched off. The same distance sweep as described above was repeated on the dark position. To ensure that no drift occurred during acquisition, follow-up curves on each of the two positions were taken. If the distance-dependent spin-polarized current curves of these two full sets of data acquisition exhibit a shift along the $z$-axis of more than 5 pm (which is more than the typical variation between the forward and backward sweep in the current), then the data was disregarded. Otherwise, the different curves at each of the two positions with opposite spin contrast were averaged, and the frequency-shift difference $\Delta f_{ex}$ was calculated. Prior to subtracting, the $\Delta f(z)$ curves were smoothed by a Savitzky-Golay filter (polynomial order: 1, frame width: 7). The magnetic exchange force $F_{ex}$ was calculated (Ref. [8]) from $\Delta f_{ex}(z)$ using formula 1. The shown magnetic exchange force curves in Fig. 4 and Fig. S6 were smoothed by Savitzky-Golay filter (polynomial order: 1, frame width: 5).

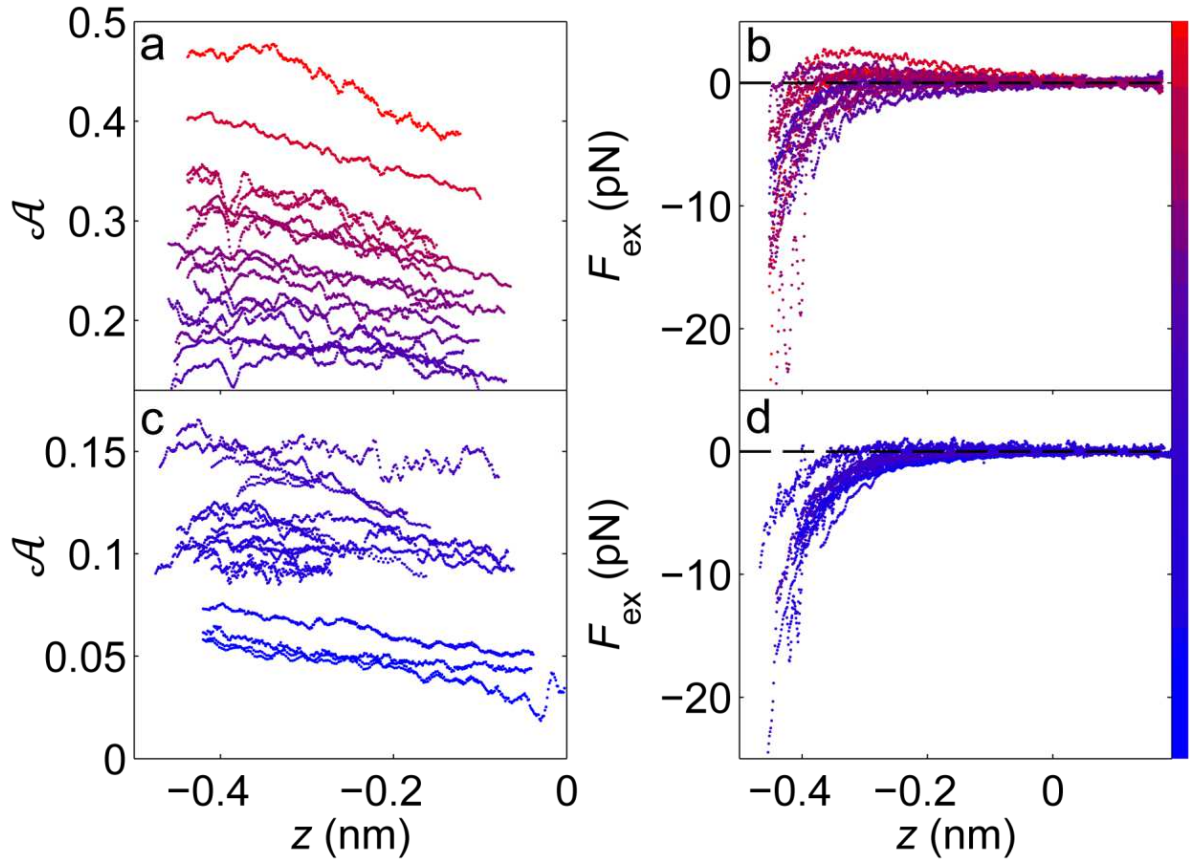

**Figure S6: Distance-dependent measurements on the fcc Fe nano-skyrmion lattice grouped by spin polarization:** (a) Spin-polarized asymmetry for mainly out-of-plane magnetized tips with $\mathcal{A}(z) \geq 0.15$, and (c) with $\mathcal{A}(z) \leq 0.15$. (b) and (d) display the corresponding exchange force $F_{ex}(z)$, respectively. The color code for all sub-figures reflects the averaged spin-polarized asymmetry $\mathcal{A}$ with blue to red representing low to high.

In order to illustrate the reproducibility and spread of our data, we show all acquired datasets in Fig. S6 for $F_{ex}(z)$ derived from $\Delta f_{ex}(z)$ as well as the spin-polarized current asymmetry ($\mathcal{A}(z)$). The data is split into two groups of larger ($\mathcal{A}(z) \geq 0.15$, (a)) and smaller ($\mathcal{A}(z) \leq 0.15$, (c)) spin-polarized current asymmetry, respectively, together with the respective magnetic exchange force $F_{ex}(z)$ ((b) and (d)). The color code for all sub-figures reflects the averaged spin-polarized asymmetry $\mathcal{A}(z)$ with blue to red representing low to high. For all shown curves, no drastic increase in the tuning-fork excitation voltage was observed, which would have indicated a non-stable tip with dissipative processes[9].

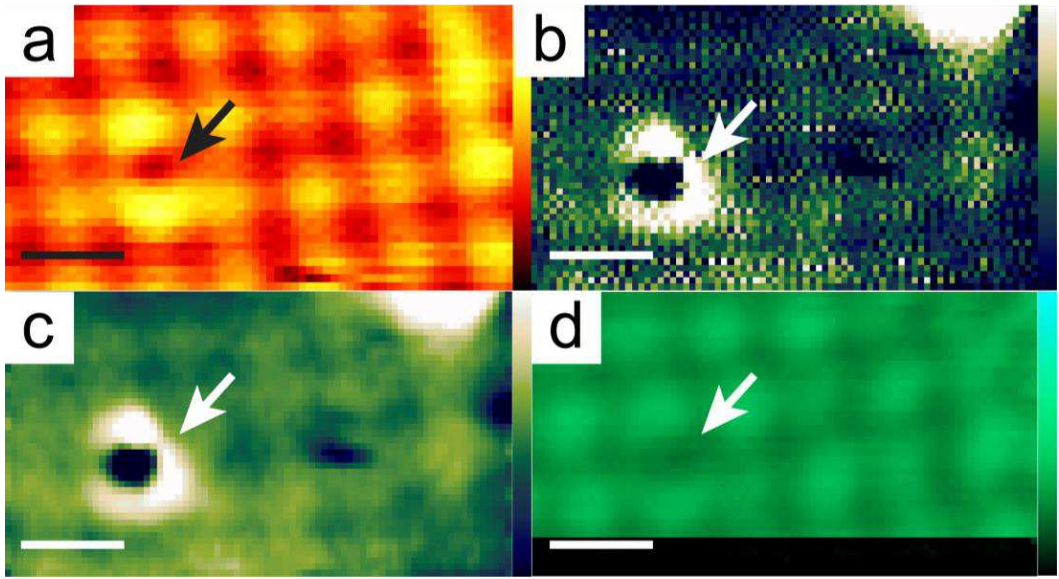

**Figure S7: MExFM image of the fcc skyrmion lattice on first layer Fe on Ir with changed voltage:** (a) SP-STM constant-current topography of the skyrmion lattice on the fcc Fe monolayer (sizes all images: 5 x 2.8 nm$^2$, $V_S = 50$ mV, $I_t = 100$ pA). (b) MExFM image of the lower part of (a), $\Delta f_{set}$: –36 Hz, $V_s = 0.0$ mV (lower part), $V_s = 0.4$ mV (upper part), $z_{mod} = 180$ pm)). (c) same image as (b), but smoothed by a 2-point Gaussian filter. (c) Simultaneously measured current map. At the scan line marked by the black arrow the voltage was increased from 0 to 0.4 mV. The arrows in the images mark the same defect. A plane has been subtracted from all images. The scale bar is 1 nm. Color scales: (a) 3.7-49 pm, (b) 18.6-22.1 pm, (c) 12.7-18.4 pm, (d) 0.0-4.4 nA.

## S8. Discussion of cross-talk

In combined STM/AFM imaging with a tuning fork, it is important to rule out potential cross-talk between the current and the frequency shift, where the tunneling current may introduce an interference with the deflection of the tuning fork. It has been suggested before to use a separate wire for the tunneling current to minimize this unwanted signal.[1,2] For our measurements we used two different types of tuning forks, as described above. For both forks, we checked for cross-talk by changing the applied bias voltage, and thus the current, while acquiring a MExFM image. Fig. S7 shows an SP-STM image (a) and an MExFM image (b,c) together with the simultaneously measured current (d) for the fcc skyrmion lattice on the Fe monolayer on Ir(111). At the horizontal position marked by the black arrow in (d), the voltage was changed from $V_S = 0.0$ mV (lower part) to $V_S = 0.4$ mV (upper part). While there is a clear signal change in the current image, no change in the MExFM corrugation is observed. Fig. S8 shows a constant-height image of the frequency shift ($\Delta f$) (a,b) and the simultaneously acquired current map. At the line indicated by the arrow, the bias voltage was changed from $V_S = 0.1$ mV (upper part) to 0 mV (lower part). Again, while there is a clear signal change in the current image, no change in the MExFM corrugation is observed. This further reveals that the AFM image at close tip-surface distances corresponds to a MExFM image where the signal is attributed to the local variations in exchange force between the magnetic tip and the surface nano-skyrmion lattice. We note that the tip used to acquire the images in Fig. S8 was one out of two tips for which the maxima in the corrugation in $\Delta f$ and in the spin-polarized current are not at the same positions (see section S6).

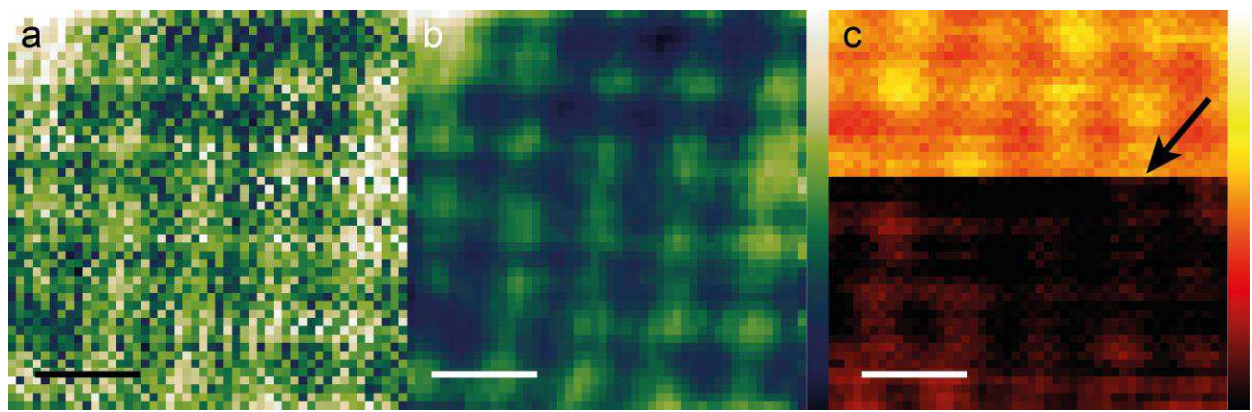

**Figure S8**: **Constant-height imaging of the fcc skyrmion lattice on the Fe monolayer on Ir(111):** (a) Frequency shift $\Delta f$ (raw data), (b) smoothed by a Gaussian filter, and (c) simultaneously acquired current (raw data). Size: 4.0 x 4.0 nm$^2$ for all images, $z_{mod} = 51$ pm, $z = -400$ pm with the current-feedback loop opened above the bright position of the skyrmion lattice at $V_s = 50$ mV and $I_t = 100$ pA. At the line indicated by the arrow, the bias voltage was changed from $V_S = 0.1$ mV (upper part) to 0 mV (lower part). There is a small voltage offset ($\approx |0.05|$ mV to $|0.1|$ mV) which is not constant in time and which could not be compensated entirely. While there is a clear change in the current, the $\Delta f$ images (a) and (b) show no change. This shows that the corrugation in the $\Delta f$ image is not influenced by any potential cross-talk with the current. The scale bar is 1 nm in all images. Color scales: (a) and (b) –34.9 to –33.6 Hz, (c) –135 pA to 290 pA.